\documentclass[12pt]{article}
\usepackage[utf8]{inputenc}
\usepackage[T1]{fontenc}
\usepackage{multirow}
\usepackage{verbatim}
\usepackage{longtable}
\usepackage{booktabs}
\usepackage{amssymb, amsmath, amsthm}
\usepackage{bbm}
\usepackage[english]{babel}
\usepackage{lmodern, csquotes, eurosym}
\usepackage[style=authoryear,texencoding=utf8,backend=biber, minnames = 1, maxnames = 4]{biblatex}
\usepackage{graphicx,subcaption}
\usepackage[colorlinks=true, citecolor = blue, linkcolor = blue]{hyperref}
\usepackage{geometry}
\usepackage{pdflscape}
\usepackage{tikz}
\usetikzlibrary{arrows}
\usetikzlibrary{calc}
\usetikzlibrary{patterns}
\usepackage{blkarray}
\usepackage{derivative}
\usepackage{nicefrac}
\usepackage{xfrac}
\usepackage{float}

\newtheorem{theorem}{Theorem}

\newtheorem{condition}{Condition}

\newtheorem{proposition}{Proposition}

\newcommand{\argmaxF}{\mathop{\mathrm{argmax}}\limits}
\renewcommand{\P}{\mathbb{P}}

\DeclareMathOperator*{\supp}{supp}

\title{
Testing Capacity-Constrained Learning\thanks{We thank Federico Echenique, Mark Dean, and Gerelt Tserenjigmid for helpful comments. We also thank the Sloan Foundation for support under the “Cognitive Economics at Work” grant.}
}
\author{Andrew Caplin\thanks{Department of Economics, New York University.}, \ Daniel Martin\thanks{Department of Economics, University of California, Santa Barbara.}, \ Philip Marx\thanks{Department of Economics, Louisiana State University.}, \\ \ Anastasiia Morozova\footnotemark[3], \ and Leshan Xu\footnotemark[3]}
\date{January 31, 2025}

\begin{document}

\maketitle

\begin{abstract}
We introduce the first general test of capacity-constrained learning models. Cognitive economic models of this type share the common feature that constraints on perception are exogenously fixed, as in the widely used fixed-capacity versions of rational inattention (\cite{sims2003implications}) and efficient coding (\cite{woodford2012prospect}). We show that choice data are consistent with capacity-constrained learning if and only if they satisfy a \emph{No Improving (Action or Attention) Switches (NIS)} condition. Based on existing experiments in which the incentives for being correct are varied, we find strong evidence that participants fail NIS for a wide range of standard perceptual tasks: identifying the proportion of ball colors, recognizing shapes, and counting the number of balls. However, we find that this is not true for all existing perceptual tasks in the literature, which offers insights into settings where we do or do not expect incentives to impact the extent of attention.
\end{abstract}

\begin{quote}
\textbf{Key words:} Cognitive economics, rational inattention, efficient coding, subjective perception, Bayesian learning, incentives, experiments  \\\smallskip
\textbf{JEL codes:} D60, D83, D91
\end{quote}

\newpage  

\section{Introduction}

Cognitive economics has emerged as a leading approach for understanding human cognition. Motivated by a long literature on subjective perception in cognitive science (e.g., \cite{weber1834pulsu}), this approach blends economic, psychology, and neuroscience methods to better understand the limits of human perception (\cite{caplin2024data}). 
One of the main divides in cognitive economics is whether to treat the bounds on perception (``perceptual capacities'' or ``perceptual constraints'') as exogenously fixed or not. For example, in the rational inattention literature, \textcite{sims2003implications} assumes the bounds on attention are fixed, and \textcite{matvejka2015rational} assumes the bounds are elastic. In the efficient coding literature, \textcite{woodford2012prospect} provides versions of optimal coding with both fixed bounds and elastic bounds.\footnote{Fixed-capacity efficient coding has been applied to risky choice (\cite{khaw2021cognitive}, \cite{frydman2022efficient}), investing (\cite{charles2024insensitive}), probability weighting (\cite{frydman2023source}), belief updating (\cite{ba2022over}), and play in games (\cite{frydman2021coordination}).}

Whether or not perceptual capacity is exogenously fixed has substantial implications for the economic impact of bounded cognition. Most importantly, if perceptual constraints are exogenous, then changes in incentives will not alter the extent of attention, just how individuals decide to allocate their scarce attention. Whether incentives alter attention has become a key question for a growing literature on the role of cognition in economics (e.g., \cite{bronchetti2023attention}).\footnote{A halfway house between these two models is one in which there learning is endogenous but bounded, as in \textcite{filiz2024information}.}

The question of whether incentives impact the extent of attention also relates to an ongoing debate about whether incentivization matters in experiments. Various reasons have been given for incentivizing experiments (e.g., \cite{plott1986rational}, \cite{smith1991rational}), but a long literature in psychology treats incentives as irrelevant to cognition, so it does not incentivize performance in experiments on perception and attention. However, this assumption has been challenged by papers showing the impact of rewards on visual perception in the psychology literature on psychometrics (e.g., \cite{pessoa2010embedding}). 

We offer a method for answering the question of whether perceptual capacity is exogenously fixed, and hence whether incentives impact the extent of attention, by providing a sharp test of capacity-constrained learning models, a class of models in which decision-makers choose among a feasible set of ways to learn. Because feasibility is an exogenous constraint, this model class covers exogenous perception constraints, including fixed-capacity versions of rational inattention (\cite{sims2003implications}) and efficient coding (\cite{woodford2012prospect}). 

We show that a single condition, \emph{No Improving (Action or Attention) Switches (NIS)}, is both necessary and sufficient for capacity-constrained learning. NIS specializes the test of costly learning from \textcite{caplin2015revealed} by replacing their \emph{No Improving Attention Cycles (NIAC)} condition and the \emph{No Improving Action Switches (NIAS)} condition of \textcite{caplin2015testable} with a single new condition, which implies both NIAS and NIAC, but is stronger than both. In words, NIS states that \emph{wholesale} switches of attention or actions, both within and across decision problems, should not improve utility.

We apply our test of capacity-constrained learning to data from three existing experiments that vary the payoffs to being correct while performing visual perception tasks. We first look at experiments 1.2 and 2.2 of \textcite{dean2023experimental}, which vary the reward to correctly guessing the state in two tasks: the number of colored balls in a visual display (1.2) and the number of correct equations (2.2).  Using the aggregated choices of participants, we find that NIAS and NIAC are satisfied for both experiments. However, NIS fails in both experiments, and the failure is statistically significant in Experiment 1.1 ($p<0.01$).%
\footnote{One possible reason that the violation of NIS in experiment 2.2 is not statistically significant is that performance is already high (over 80\%) at very low incentives, which leaves little room for improvement with incentives. Another possible reason is that there are only two incentive levels, so there are few inequalities to check.}
Importantly, NIS fails in a systematic way in both experiments: switches in attention in the direction of higher incentives always lead to improvements in expected utility. 

Next, we reexamine the data from the experiment of \textcite{caplin2020rational}, which varies the payoff for correctly guessing which shape appears the most frequently. We find that NIAS is satisfied for all four difficulty levels of the task, but that NIS is not satisfied for any of the difficulty levels ($p<0.01$ for all four).

Finally, we re-analyze the data from the experiment of \textcite{dewan2020estimating}, which varies the payoffs at two possible prize sizes (\$10 and \$20). For the first task, correctly guessing the number of balls in a display, we again find that NIAS and NIAC are satisfied for both prize sizes, but NIS fails for both ($p<0.01$ for each prize size).
As in the other experiments, we find that for a given prize size, switches of attention in the direction of higher incentives always lead to improvements in expected utility.

However, for the second task of \textcite{dewan2020estimating}, which requires participants to judge the degree of an angle, we do not find strong evidence that NIS fails for either of the prize sizes ($p=0.45$ and $p=0.72$). In addition, unlike the other experiments, switches of attention in the direction of higher incentives (for a given prize size) often do not lead to improvements in expected utility. 

These results suggest that certain task factors might dictate the suitability of capacity-constrained learning. Specifically, this model class might be more appropriate for tasks that are not responsive to additional effort.\footnote{The relationship between incentives and the effort required to complete judgment tasks is explored in \textcite{camerer1999effects}.} This can occur because (1) even at low incentives, an individual chooses to exert enough effort to reach their perceptual limit, and (2) additional effort would not meaningfully improve perception beyond this limit. This combination of forces causes individuals to reach the same level of ``irreducible uncertainty'' at all levels of incentives. Another possibility is that uncertainty could actually be reduced further, but individuals choose not to spend the additional effort required to reduce it further at standard incentive levels.

Our paper contributes to a growing literature on testing economic models of cognition on human decision makers (e.g., \cite{caplin2020rational,de2020communication,dewan2020estimating,dean2023experimental,almog2024rational,almog2024ai}).
Our primary contribution to this literature is to examine the consistency of human choices with capacity-constrained learning models.

Our test also has implications for non-human learners. \textcite{caplin2024modeling} apply our test to the predictions of a state-of-the-art machine learning algorithm that is trained to identify pneumonia from chest x-rays. In the context of machine learning, capacity-constrained learning aligns with standard notions of how a machine learns, as this model class can be interpreted as the machine choosing among a feasible set of mathematical operations to best match the incentives provided by its loss function. Similarly to what we find in many existing experiments with humans, \textcite{caplin2024modeling} find that the predictions of the machine learning algorithm violate NIS but do not violate NIAS or NIAC.

The rest of the paper is as follows. Section \ref{sec:test} introduces NIS, our test of capacity-constrained learning, and Section \ref{sec:existing} describes how NIS relates to two existing tests of Bayesian learning (NIAS and NIAC). Section \ref{sec:representation} formalizes capacity-constrained learning models and their relationship to NIS. Section \ref{sec:data} proposes approaches to measuring the extent of NIS violations and the statistical testing of NIS. Section \ref{sec:results_existing} provides the results of testing NIS using the data from existing experiments. 

\section{Our Test of Capacity-Constrained Learning}\label{sec:test}

\subsection{Preliminaries}
\label{sec:preliminaries}

There is a finite set of possible states of the world $\omega\in\Omega$ distributed according to an objective prior,%
\footnote{The tests in this paper can be readily extended to settings where the objective prior varies, but it is not necessary for testing capacity-constrained learning.}
a finite global set of actions $a \in \mathcal{A}$ with $| \mathcal{A} | \geq 2$, and a finite global set $x \in X$ of prize specifications, with realizations $x(a,\omega)$ that depend on the realized state and action in a way known by the researcher. A decision problem $i \in D$ consists of an action set $A_i \subseteq \mathcal{A}$ and a prize specification $x_i \in X$.%
\footnote{
Our specification is similar to that of \textcite{dean2023experimental}, who vary both action sets and prizes in their experiments.
To reconcile with \textcite{caplin2015revealed}, note that a set of actions $a$ and prizes $x$ can always be reduced to a set of actions only by redefining $a \leftarrow (a,x)$.
}

For example, Experiment 1.2 of \textcite{dean2023experimental} (DN23 hereafter) has two equally likely states: out of the 100 blue and red balls presented on the screen, there can either be 49 red balls or 51 red balls (labeled states $\omega_1$ and $\omega_2$, respectively).
The global set of actions $\mathcal{A}=\{a_1,a_2\}$ is common across decision problems, and individuals are incentivized to match actions and states --- choosing $a_1$ in state $\omega_1$ and $a_2$ in state $\omega_2$ --- according to a varying prize specification such that a correct guess leads to either 5, 40, 70, or 95 \emph{probability points} (the probability of receiving a fixed cash amount) while an incorrect guess yields 0 probability points. 
To summarize, the set of decision problems $i \in \{1,2,3,4 \}$ consists of common states and actions yielding varying probability point prizes, which are known to correspond to actions and states as follows:

\begin{equation*}
\Omega = \{ \omega_1, \omega_2 \}, \quad
A_i = \{ a_1, a_2 \}, \quad x_i (a,\omega)=
\begin{blockarray}{ccc}
\omega_1 & \omega_2 &\\
\begin{block}{(cc)c}
   p_i & 0 & a_1 \\
   0 & p_i & a_2\\
\end{block}
\end{blockarray}
\quad \text{where $p_i = \begin{cases} 
5 & \text{for $i = 1$}, \\
40 & \text{for $i = 2$}, \\
70 & \text{for $i = 3$}, \\
95 & \text{for $i = 4$}.
\end{cases}$}
\end{equation*}

The decision-maker (DM) is assumed to have a utility function $u$ over prizes. In all of the experiments we study, the prizes are a number of probability points, and under the assumption of expected utility, the utility of a probability point prize is linear in the number of probability points. If we assume that receiving cash yields more utility than not receiving it, then the utility of the prize is also strictly increasing in the number of probability points. Without loss of generality, we can normalize the utility of receiving the cash prize to 100 and not receiving the cash to 0, in which case the utility of the probability point prize is point identified: the utility of $p$ probability points is exactly $p$. 

As in \textcite{caplin2015testable}, the data relevant to assessing what a DM with private information choosing in decision problem $A$ is state dependent stochastic choice (SDSC) data. This specifies, for each decision problem $i \in D$, the joint distribution of actions and states $P_i (a,\omega)$ for all $a\in A_i$ and $\omega\in\Omega$. We denote as $P$ the collection of $P_{i}$ for every decision problem $i\in D$.

For example, in Experiment 1.2 of DN23, the set of aggregate (across-subject) joint distributions $P$ is:
\begin{center}
$P_{1}=
\begin{blockarray}{ccc}
\omega_1 & \omega_2 &\\
\begin{block}{(cc)c}
   0.37 & 0.20 & a_1 \\
   0.13 & 0.30 & a_2\\
\end{block}
\end{blockarray}
\quad \quad
\text{} 
\quad \quad
P_{2}=
\begin{blockarray}{ccc}
\omega_1 & \omega_2 &\\
\begin{block}{(cc)c}
   0.38 & 0.17 & a_1 \\
   0.12 & 0.33 & a_2\\
\end{block}
\end{blockarray}\quad \quad
\text{}$
\end{center}

\begin{center}
$P_{3}=
\begin{blockarray}{ccc}
\omega_1 & \omega_2 &\\
\begin{block}{(cc)c}
   0.39 & 0.17 & a_1 \\
   0.11 & 0.33 & a_2\\
\end{block}
\end{blockarray}
\quad \quad
\text{} 
\quad \quad
P_{4}=
\begin{blockarray}{ccc}
\omega_1 & \omega_2 &\\
\begin{block}{(cc)c}
   0.39 & 0.14 & a_1 \\
   0.11 & 0.36 & a_2\\
\end{block}
\end{blockarray}\quad \quad
\text{}$
\end{center}

\subsection{NIS: Capacity-Constrained Learning}

Consider an econometrician who wants to determine if a decision maker chose in line with capacity-constrained learning, but only 
observes state-dependent stochastic choices (SDSC), 
rather than the information actually obtained by the decision maker. The condition for a collection of SDSC data $P$ to be consistent with capacity-constrained learning is given below.%
\footnote{A precise definition of capacity-constrained learning, and our representation result is in Section \ref{sec:representation}.}

\begin{condition}[No Improving (Action or Attention) Switches (\textbf{NIS})] 
Utility function $u$ satisfies NIS for $P$ if and only if for any set of decision problems $i,j \in D$, 
\begin{eqnarray}
\label{eqn:NIS}
\sum_{a\in A_i} \sum_{\omega\in\Omega} P_{i}(a,\omega) u(x_i(a,\omega)) \geq 
\sum_{a\in A_j} \max_{\hat{a}\in A_i} \sum_{\omega\in\Omega} P_{j}(a,\omega) u(x_i (\hat{a},\omega))
\end{eqnarray}
\end{condition}

In words, this condition states that \emph{wholesale} switches of attention and actions should not improve utility according to the baseline prize specification. NIS involves both checks between decision problems (when $i \neq j$), where both actions and choice probabilities can change, and checks ``within'' each decision problem (when $i = j$), where only actions can change. Thus, NIS involves checking $N^2+N$ inequalities.

\subsubsection{Example: DN23}
\label{sec:dn23nis}
To illustrate NIS we turn to the decision problems, prizes, and aggregate SDSC data from Experiment 1.2 of DN23, which are given in Section \ref{sec:preliminaries}. Restricting for simplicity to just the first two decision problems ($1,2$) in which the number of probability points was 5 and 40 respectively, the data $P$ is:
\begin{center}
$P_{1}=
\begin{blockarray}{ccc}
\omega_1 & \omega_2 &\\
\begin{block}{(cc)c}
   0.37 & 0.20 & a_1 \\
   0.13 & 0.30 & a_2\\
\end{block}
\end{blockarray}
\quad \quad
\text{} 
\quad \quad
P_{2}=
\begin{blockarray}{ccc}
\omega_1 & \omega_2 &\\
\begin{block}{(cc)c}
   0.38 & 0.17 & a_1 \\
   0.12 & 0.33 & a_2 \\
\end{block}
\end{blockarray}\quad \quad
\text{}$
\end{center}
The NIS condition makes four comparisons: switches of actions within (decision) problem $1$, switches of actions within problem $2$, switches of choice probabilities and actions from problem $1$ to $2$, and switches of choice probabilities and actions from problem $2$ to $1$. In this case, switches of choice probabilities from problem $1$ to $2$ generate a violation of NIS, which we will show below.

The value of the data in decision problem $1$ is
\begin{eqnarray*}
& \underbrace{P_{1}(a_1, \omega_1) u(x_1 (a_1, \omega_1))}_{0.37 \times 5} + \underbrace{P_{1}(a_1, \omega_2)  u(x_1 (a_1, \omega_2))}_{0.20 \times 0} +  \underbrace{P_{1}(a_2, \omega_1) u(x_1(a_2, \omega_1))}_{0.13 \times 0} + \underbrace{P_{1}(a_2, \omega_2) u(x_1 (a_2, \omega_2))}_{0.30 \times 5} 
\\
&= 3.34.
\end{eqnarray*}
Switching to the data  in problem 2 while keeping the optimal actions and prizes in problem $1$ results in 
\begin{eqnarray*}
& \underbrace{P_{2}(a_1, \omega_1) u(x_1 (a_1, \omega_1))}_{0.38 \times 5} + \underbrace{P_{2}(a_1, \omega_2)  u(x_1 (a_1, \omega_2))}_{0.17 \times 0} +  \underbrace{P_{2}(a_2, \omega_1) u(x_1(a_2, \omega_1))}_{0.12 \times 0} + \underbrace{P_{2}(a_2, \omega_2) u(x_1 (a_2, \omega_2))}_{0.33 \times 5} 
\\
&= 3.55.
\end{eqnarray*}
Because the optimal actions remain the same and choice probabilities on the diagonal increase, switching results in a net gain of utility ($0.21$). Thus, there would have been an improving switch in utility by using the implied learning in decision problem $A_2$ in decision problem $A_1$. As a result, NIS is not satisfied for this data.

\section{Relationship to Existing Tests}\label{sec:existing}

We now describe NIAS and NIAC, two existing tests of Bayesian learning, and explain their relationship to NIS. Overall NIS implies both NIAS and NIAC, but neither implies NIS, so NIS is strictly stronger than both. It is also strictly stronger than their combination. Throughout, we will continue to illustrate the tests using aggregate choice data from Experiment 1.2 of DN23.

\subsection{NIAS: Optimal Actions Given Information}
\label{sec:nias}

The key assumptions behind many, if not most, models of attention and perception is that decision makers update beliefs correctly (using Bayes' Rule) and maximize expected utility given some unobservable learning about the state. In other words, they choose optimally given the (private) information that they have acquired.

\textcite{caplin2015testable} show that, for utility function $u:X \rightarrow \mathbb{R}$ and some private information, SDSC data is consistent with choosing optimally given that information if and only if the \emph{No Improving Action Switches (NIAS)} condition is satisfied. This condition requires that no wholesale switch in actions improves utility. Formally, utility function $u$ satisfies the NIAS condition with respect to $P$ if for every decision problem $i\in D$ and actions $a, \hat{a} \in A_i$,
\begin{equation}\label{eqn:NIAS}
\sum_{\omega \in \Omega }P_i(a,\omega)u(x_i(a,\omega))\geq
\sum_{\omega \in \Omega }P_i(a,\omega)u(x_i(\hat{a},\omega)), 
\end{equation}
NIS implies NIAS because NIAS is satisfied if and only if DM satisfies NIS within each decision problem. That is, when $i=j$, NIS (\ref{eqn:NIS}) becomes:
\begin{equation}
\label{eq:NIAS-equiv}
\sum_{a\in A_i} \sum_{\omega\in\Omega} P_{i}(a,\omega) u(x_i(a,\omega)) = 
\sum_{a\in A_i} \max_{\hat{a}\in A_i} \sum_{\omega\in\Omega} P_{i}(a,\omega) u(x_i(\hat{a},\omega)),
\end{equation}
which is satisfied if and only if NIAS holds.

\subsubsection{Example from DN23}

To illustrate NIAS we turn to the decision problems, prizes, and aggregate SDSC data from Experiment 1.2 of DN23, which are given in Section \ref{sec:preliminaries}. Within each decision problem $i$ of Experiment 1.2 of DN23, 
the NIAS condition reduces to
\begin{eqnarray*}
P_{i}(a_1,\omega_1) \geq
P_{i}(a_1,\omega_2), \\
P_{i}(a_2,\omega_2) \geq
P_{i}(a_2,\omega_1),
\end{eqnarray*}
which is clearly satisfied.
Note that this condition would hold for any SDSC data in which the diagonal entry of the matrix is larger than the entries in each row, consistent with the incentive to match the action to the state. 

\subsection{NIAC: Optimal Costly Learning}

Models of rational inattention also assume that decision-makers optimally acquire their (private) information based on the cost of that information and the expected utility of choosing in line with that information. \textcite{caplin2015revealed} characterize the optimal choice of costly information by pairing NIAS with the \emph{No Improving Action Switches (NIAC)} condition, which states that utility cannot be improved by rotating choice probabilities between decision problems.

\begin{condition}
[No Improving Attention Cycles (\textbf{NIAC})] Utility function $u$ satisfies NIAC for $P$ if for any sequence of decision problems $1, 2, \dots, J \in D$ with convention $J+1 = 1$,
\begin{eqnarray*}
\sum_{j=1}^{J}\left(\sum_{a\in A_{j}} \max_{\hat{a}\in A_{j}} \sum_{\omega\in\Omega} P_{j}(a,\omega) u(x_j(\hat{a},\omega))\right) \geq \\
\sum_{j=1}^{J}\left(\sum_{a\in A_{j+1}} \max_{\hat{a}\in A_{j}} \sum_{\omega\in\Omega} P_{j+1}(a,\omega) u(x_j(\hat{a},\omega))\right)
\end{eqnarray*}
\end{condition}
Note that together, NIAS and NIAC imply that 
\begin{eqnarray*}
\sum_{j=1}^{J}\left(\sum_{a\in A_{j}}  \sum_{\omega\in\Omega} P_{j}(a,\omega) u(x_j(a,\omega))\right) \geq \\
\sum_{j=1}^{J}\left(\sum_{a\in A_{j+1}} \max_{\hat{a}\in A_{j}} \sum_{\omega\in\Omega} P_{j+1}(a,\omega) u(x_j(\hat{a},\omega))\right)
\end{eqnarray*}
If we consider every binary comparison instead of every cycle, then NIAC reduces to NIS. Thus, NIS implies NIAC. Intuitively, if every binary comparison is non-improving, then all cycles must be non-improving.

\subsubsection{Example from DN23}

To illustrate NIAC we again turn to the decision problems, prizes, and aggregate SDSC data from Experiment 1.2 of DN23, which are given in Section \ref{sec:preliminaries}. If Experiment 1.2 of DN23 consisted of just decision problems 1 and 2, in which the number of probability points was 5 and 40 respectively, the NIAC condition would boil down to determining whether rotating the choice probabilities between these decision problems improves utility on net:

\begin{center}
$P_{1}=
\begin{blockarray}{ccc}
\omega_1 & \omega_2 &\\
\begin{block}{(cc)c}
   0.37 & 0.20 & a_1 \\
   0.13 & 0.30 & a_2\\
\end{block}
\end{blockarray}
\quad \quad
\text{} 
\quad \quad
P_{2}=
\begin{blockarray}{ccc}
\omega_1 & \omega_2 &\\
\begin{block}{(cc)c}
   0.38 & 0.17 & a_1 \\
   0.12 & 0.33 & a_2\\
\end{block}
\end{blockarray}\quad \quad
\text{}$
\end{center}
Under the maintained assumptions of expected utility monotonically increasing in prizes and normalized as discussed in Section \ref{sec:preliminaries}, the value of the revealed signal structure in decision problem $2$ is
\begin{eqnarray*}
& \underbrace{P_{2}(a_1, \omega_1) u(x_2 (a_1, \omega_1))}_{0.38 \times 40} + \underbrace{P_{2}(a_1, \omega_2)  u(x_2 (a_1, \omega_2))}_{0.17 \times 0} +  \underbrace{P_{2}(a_2, \omega_1) u(x_2(a_2, \omega_1))}_{0.12 \times 0} + \underbrace{P_{2}(a_2, \omega_2) u(x_2 (a_2, \omega_2))}_{0.33 \times 40} 
\\
&= 28.4.
\end{eqnarray*}
Switching to the signal structure revealed in problem 1 requires re-optimizing for prizes in problem 2:
\begin{eqnarray*}
    & \max_{a \in \{a_1, a_2\}}[P_{1}(a_1, \omega_1)u(x_2 (a, \omega_1))+P_{1}(a_1,\omega_2)u(x_2 (a, \omega_2))]\\
    + & \max_{a \in \{a_1, a_2\}}[P_{1}(a_2, \omega_1)u(x_2 (a, \omega_1))+P_{1}(a_2, \omega_2)u(x_2 (a, \omega_2))] 
\end{eqnarray*}
However, given that NIAS is satisifed in each decision problem, the optimal actions remain the same and so the preceding term equals: 
\begin{eqnarray*}
& \underbrace{P_{1}(a_1, \omega_1) u(x_2 (a_1, \omega_1))}_{0.37 \times 40} + \underbrace{P_{1}(a_1, \omega_2)  u(x_2 (a_1, \omega_2))}_{0.20 \times 0} +  \underbrace{P_{1}(a_2, \omega_1) u(x_2(a_2, \omega_1))}_{0.13 \times 0} + \underbrace{P_{1}(a_2, \omega_2) u(x_2 (a_2, \omega_2))}_{0.30 \times 40} 
\\
&= 26.8
\end{eqnarray*}
Because the optimal actions remain the same and choice probabilities on the diagonal decrease, this switch in learning results in a net loss ($\approx 26.8 - 28.4 = -1.6$). 

Recall from Section \ref{sec:dn23nis} that the value of the revealed signal structure in (decision) problem 1 is $3.34$, whereas the value of the revealed signal structure in problem 2 but using the prize specification of problem 1 is $3.55$. 
Switching from signal structure $P_1$ to $P_2$ in problem 1 results in a net gain of utility ($\approx 3.55 - 3.34 = 0.21$). As discussed in Section \ref{sec:dn23nis}, this produces a violation of the NIS condition, but because there is no overall improvement ($-1.6+0.21<0$), the NIAC condition is satisfied.

\section{Characterizing Capacity-Constrained Learning}\label{sec:representation}


Our main goal is to characterize which learning strategies could have rationally generated the observed data $P$ under the model of capacity-constrained learning.
The decision models that we consider involve rational Bayesian learning and subsequent optimal choice. Our notation broadly follows \textcite{caplin2022bayesianlearning}, henceforth CMM. As in \textcite{kamenica2011bayesian}, we specify conceivable learning as a Bayes consistent distribution $Q$ of posteriors $\gamma\in\Delta(\Omega)$ with finite support $\Gamma (Q) \equiv \supp Q$.   
We refer to such distributions of posteriors as information structures, with their set given by:
\[
\mathcal{Q}\equiv\{Q\in\Delta(\Delta(\Omega))\text{ with }| \Gamma (Q) |<\infty \text{ and } \sum_{\gamma\in\Gamma (Q)}\gamma Q(\gamma)=\mu\}.
\]
Once learning has taken place, the DM selects a mixed strategy over actions as a function of the posterior, $q(a|\gamma)\in\Delta(A)$. 
%
Define $P_{(Q,q)}$ as the hypothetical SDSC that any strategy $(Q,q) $ would generate,
\begin{equation}
\label{eq:PQq}
    P_{(Q,q)}(a,\omega) \equiv \sum_{\gamma\in \Gamma(Q)}q(a|\gamma)Q(\gamma)\gamma(\omega).
\end{equation}
Then a strategy $(Q,q)$ generates the data $P$ if:
\begin{equation}
    \label{eq:Qq-generate}
    P_{(Q,q)} = P
\end{equation}
We will say that such a strategy \emph{rationalizes} the data if it furthermore arises from optimal choice.

We model the DM's optimization problem in two stages, which we solve using backward induction.
In the second stage, given an information structure $Q$ and decision problem $A$, the DM chooses an action strategy to maximize expected utility.
Specifically, given a prize specification $x$ and a posterior $\gamma$, define the posterior expected utility as:%
\footnote{
We depart slightly from \textcite{caplin2022bayesianlearning} by parametrizing expected utility as a function of prizes rather than Bernoulli utility $u$. 
This is because throughout the experiments in this paper, prizes are varied whereas a subject's utility of money is held fixed. 
}
\begin{equation}
    \label{eq:posterior-U}
    U(a | \gamma, x) \equiv \sum_{\omega \in \Omega} \gamma (\omega) u(x(a,\omega)),
\end{equation}
and define the gross expected utility of strategy $(Q,q)$ given $u$ as:
\[
g(Q,q|x) \equiv \sum_{\gamma \in \Gamma(Q)}\sum_{a\in \mathcal{A}} Q(\gamma )q(a|\gamma ) U(a | \gamma, x).
\]
Then in the second stage as a function of learning $Q$, the DM chooses an action strategy
to solve:
\begin{equation}
\label{eq:eu-max}
    \argmaxF_{q: \Gamma (Q) \to \Delta (A)} \, g(Q,q|x) 
\end{equation}
In what follows, it will also be useful to define the resulting indirect expected utility of an information structure $Q$ in decision problem $A$ given utility function $u$ as:
\begin{equation}
\label{eq:G}
G(Q | A,x) \equiv \max_{q: \Gamma (Q) \to \Delta (A)} \, g(Q,q|x)
\end{equation}
In the first stage, the DM chooses an information structure to maximize this indirect expected utility subject to a feasibility constraint $\mathcal{Q^*} \subset \mathcal{Q}$. That is, the DM chooses a learning strategy to solve: 
\begin{equation}
\label{eq:costly-learning}
     \argmaxF_{Q \in \mathcal{Q^*}} \, G(Q | A,x)
\end{equation}

\subsection{Example: Fixed-Capacity Rational Inattention}

In \textcite{sims2003implications}, the DM solves:
\begin{equation*}
    \max_{P}\sum_{a\in A}\sum_{\omega\in\Omega} \mu(\omega) P(a|\omega) u(x(a,\omega)) 
\end{equation*}
subject to:
\begin{align*}
\forall a \in A, \omega \in \Omega: P(a | \omega) \in [0,1], \quad \forall \, \omega \in \Omega: \sum_{a\in A}P(a|\omega)&=1, \\
\sum_{a \in A} \sum_{\omega \in \Omega} \mu(\omega) P(a|\omega)\log \left( \frac{P(a|\omega)}{\sum_{\omega \in \Omega} \mu (\omega) P(a | \omega)} \right)  &\leq C
\end{align*}
where $C$ is the Shannon capacity.

Letting $\mathcal{P}_\mu$ denote the set of \textit{joint} distributions $P$ over $A \times \Omega$ with marginal distribution $\sum_{a \in A} P(a, \omega) = \mu (\omega)$ for each $\omega \in \Omega$, we can equivalently write the problem as:
\begin{equation}
\label{eq:sims-joint}  
    \max_{P \in \mathcal{P}_\mu } \, \sum_{a\in A}\sum_{\omega\in\Omega} P(a, \omega) u(x(a,\omega)) 
\end{equation}
subject to a Shannon capacity constraint on mutual information: 
\begin{align}
\label{eq:mutual-info} 
\sum_{a \in A} \sum_{\omega \in \Omega} P(a, \omega) \log \left( \frac{P(a, \omega)}{P (a) \mu (\omega)} \right)  &\leq C
\end{align}
where $P(a) \equiv \sum_{\omega \in \Omega} P(a,\omega)$ and $\mu (\omega)$ denote marginal probabilities over action $a$ and state $\omega$, respectively.
It is now without loss of generality to restrict to the subset $\mathcal{P}_\mu^*$ of joint distributions $P \in \mathcal{P}_\mu$ that additionally satisfy NIAS \eqref{eqn:NIAS},%
\footnote{Note that $P$ here are choice variables, rather than observed data as in \textcite{caplin2015testable}, but the NIAS condition can be defined identically.}
since wholesale action switches can only relax the constraint \eqref{eq:mutual-info}, and therefore distributions $P$ violating NIAS cannot be optimal in \eqref{eq:sims-joint} subject to \eqref{eq:mutual-info}. 
For each $P \in \mathcal{P}_\mu^*$ satisfying NIAS, \textcite{caplin2015testable} guarantee the existence of a revealed experiment $Q^P$ and action strategy $q^P: \Gamma (Q^*) \to \Delta (A)$ satisfying expected utility maximization \eqref{eq:eu-max}.
Namely, for any action $a\in A$ that is chosen, $P(a)>0$, define revealed posterior $\gamma^{a}$ by:%
\begin{equation}
\label{eq:post-revealed}
\gamma^{a}(\omega )=\frac{P(a,\omega )}{P(a)};
\end{equation}%
define the 
revealed experiment $Q^P$ by:
\begin{equation}
\label{eq:exp-revealed}
Q^P(\gamma ) = \sum_{\{a\in A|\gamma^{a}=\gamma \}}P(a); 
\end{equation}
and define the action strategy by:
\begin{equation}
\label{eq:action-revealed}
q^P (a | \gamma) =\left\{ 
\begin{array}{cc}
\frac{P(a)}{Q^P(\gamma )} & \text{if }\gamma^{a}=\gamma 
\text{;} \\ 
0 & \text{if }\gamma^{a}\neq \gamma.%
\end{array}%
\right.
\end{equation} 
Taking the feasible set $\mathcal{Q}^*$ to be the set of revealed experiments associated with the set of distributions $P_\mu^*$ yields the fixed-capacity model as an instance of our capacity-constrained model.

\subsection{Example: Efficient Coding}

In the fixed-capacity models of \textcite{woodford2012prospect}, \textcite{khaw2021cognitive}, and \textcite{frydman2022efficient}, the state $\omega $ is a vector of length $N$ in a bounded subset $\Omega \subset \mathbb{R}^N$, with attributes $\omega_n$ indexed by $n$. The action $a \in A$ is also an element of a bounded subset in $\mathbb{R}^N$ with attributes $a_n$. 
For consistency with our approach and notation, we restrict to the case where the state and action spaces are furthermore finite.
The DM solves:%
\footnote{In \textcite{woodford2012prospect} and \textcite{khaw2021cognitive}, the DM's objective is to minimize the expected MSE, that is $u(a,\omega)= -(a-\omega)^2$. In \textcite{frydman2022efficient}, the DM's objective is to maximize the expected financial gain of a lottery choice question, so 
$u(a,\omega) = p\cdot\omega_X \cdot \mathbbm{1}\{c(a) = \text{Risk}\} + \omega_C \cdot \mathbbm{1}\{c(a) = \text{Certain}\}
$, with choice function
\begin{equation*}
\begin{aligned}
c(\cdot)\colon \mathcal{A}\mapsto\{\text{Risk}, \text{Certain}\} = \arg\max_{c(\cdot)} \biggl(&\iint p\cdot\omega_X \cdot \mathbbm{1}\{c(a) = \text{Risk}\} P(a|\omega) \cdot \mu(\omega),da,d\omega \\ 
&+ \iint \omega_C \cdot \mathbbm{1}\{c(a) = \text{Certain}\} \cdot P(a|\omega) \cdot \mu(\omega),da,d\omega\biggr)
\end{aligned}
\end{equation*}
}
\begin{equation} \label{eq:Woodford}
    \max_{P} \sum_{a \in A} \sum_{\omega \in \Omega}  \mu(\omega) P(a|\omega)  u(a, \omega)
\end{equation}
where (as in the previous example) $P$ is a state-conditional action probability subject to the following additional constraints. 
Since all attributes are assumed independent, $P(a|\omega)=\prod_{n=1}^N P(a_n|\omega_n)$, with $P(a_n|\omega_n)$ being a distribution over $a_n$ conditioning on $\omega_n$. 
Let $\pi_n(\cdot)$ denote a marginal distribution over state attributes $\omega_n$.
Then the efficient coding constraint is: 
\begin{equation}\label{eq:WoodfordConstrain}
    \sum_{n}\Bigg[\max_{\pi_n}\Big\{
    \sum_{a_n} \sum_{\omega_n}    \pi_n(\omega_n) P(a_n|\omega_n) 
    \log\Big[\frac{P(a_n|\omega_n)}{ \sum_{\omega_n} \pi_n (\omega_n) P(a_n | \omega_n)}
    \Big]
    \Big\}\Bigg]\leq C.
\end{equation}
As in the preceding example, we can restrict to conditional action probabilities $P$ for which the induced joint distribution defined by $P_\mu (a,\omega) \equiv \mu (\omega) P(a | \omega)$ satisfies NIAS \eqref{eqn:NIAS}, since wholesale action switches can only relax the efficient coding constraint \eqref{eq:WoodfordConstrain}.
As before, we can associate each such (conditional) probability $P$ through its induced joint distribution $P_\mu$ to a revealed experiment and action strategy $(Q^P,q^P)$ satisfying expected utility maximization \eqref{eq:eu-max}. 
Taking these associated revealed experiments $Q^P$ as the feasible set of information structures yields the efficient coding model as an instance of our capacity-constrained model.

\subsection{Representation Theorem}\label{sec:theorem}

We say that a collection of SDSC data sets $P = (P_i)_{i \in D}$ has a capacity-constrained representation if there exists a set of decision-problem-dependent strategies $(Q_i, q_i)$  for each $i \in D$ solving expected utility maximization \eqref{eq:eu-max} and optimal learning \eqref{eq:costly-learning}, given action set $A_i$, prize specification $x_i$, and a common feasible set of learning $\mathcal{Q}^*$.


\begin{theorem}
\label{thm:ccr}
    Data set $P$ has a capacity-constrained learning representation if and only if P satisfies NIS. 
\end{theorem}

\noindent
The proof is relegated to Appendix \ref{apx:proof}. Intuitively, the proof consists of two steps. The first step is to show that we can restrict the search for a capacity-constrained representation (CCR) to \textit{revealed} experiments $Q_i^P$, defined in \eqref{eq:exp-revealed}, for each decision problem $i$ and corresponding SDSC $P_i$. 
More specifically, then, the first step is to conclude that the existence of any CCR implies existence with feasible learning set $\mathcal{Q}^*\equiv\cup_i Q_i^P$ and the condition:
\begin{equation}\label{eqn:revealed}
  G(Q_i^P|A_i,x_i)\geq G(Q_j^P|A_i,x_i),
\end{equation}
for all pairs of decision problems $i,j \in D$.
The second step is to establish equivalence of the preceding condition \eqref{eqn:revealed} with the NIS condition \eqref{eqn:NIS}.

\section{Taking NIS to Data}\label{sec:data}

For the experiments that we study, in which incentives are increasing, NIS makes the stringent prediction that the percentage correct has to be identical across treatments.
Thus, it is important to establish by how much NIS would fail. We address this issue in the context of NIS in two ways. First, we propose two measures of how far away a set of choice data are from satisfying NIS. Second, we develop a method for statistically testing NIS.

\subsection{Improvability Indices}\label{sec:improv}

A known challenge in testing axioms with choice data is that axioms are either satisfied or not, which is fairly stark.  Instead, when they fail it might be useful to know the extent of the failure.

In the same spirit as a literature that measures the extent of violations of rationality for deterministic models of decision-making (\cite{afriat1973system,varian1991goodness,echenique2011money,apesteguia2015measure,dean2016measuring}), we propose two measures for the extent of violations of rationality for stochastic models of decision-making.%
\footnote{
Another approach to measuring the extent of violations of rationality for stochastic choice data is proposed by \textcite{ok2023measuring}, who
measure deviations from (possibly incomplete) preference maximization.
}
The measures are stated here for NIS, but could be extended to cover other conditions, such as NIAS and NIAC.

The first measure we propose is the \emph{Improvability Difference Index (IDI)}. This index indicates the largest difference between the actual expected utility obtained in a decision problem and what could be achieved by switching actions and attention. To aid comparability across decision problems, it is a normalized by the maximum achievable expected utility within a decision problem, so takes a value between 0 and 1.%
\footnote{If we interpret the expected utility gain from switching to the learning from another decision problem as something that a third party could extract, then IDI is in a similar spirit at the Money Pump Index (\cite{echenique2011money}), especially when adapted to the cyclical switches of NIAC.}

Technically, IDI is the larger of 0 and the (normalized) difference between the expected utility under the chosen attention in a decision problem (the left-hand side of the NIS condition \ref{eqn:NIS}) and the expected utility from using the attention chosen in a different decision problem (the right-hand side of the NIS condition  \ref{eqn:NIS}):
\begin{eqnarray}\label{eqn:IDI}
\max_{i,j \in D}{
    \frac{
        \sum_{a\in A_j} \max_{\hat{a}\in A_i} \sum_{\omega\in\Omega} P_{j}(a,\omega) u(x_i(\hat{a},\omega))-\sum_{a\in A_i} \sum_{\omega\in\Omega} P_{i}(a,\omega) u(x_i(a,\omega))
    }{ 
        \sum_{\omega\in\Omega} \max_{\hat{a}\in A_i} \mu(\omega) u(x_i(\hat{a},\omega))}
    }
\end{eqnarray}
NIS passes if and only if IDI is 0, and higher values of IDI indicate that NIS is violated by more of the maximum expected utility.

Second, we propose the \emph{Improvability Efficiency Index (IEI)}, which is the fraction of expected utility that needs to be shaved from the right-hand side of the NIS condition (the expected utility from using the attention chosen in a different decision problem) to make it smaller than the left hand side of the NIS condition (expected utility obtained in a decision problem). Technically, we calculate IEI by finding the largest value of $\epsilon\in[0,1]$ such that the following inequality holds for all sets of decision problems $A_i,A_j \in D$:
\begin{eqnarray}\label{eqn:IEI}
\sum_{a\in A_i} \sum_{\omega\in\Omega} P_{i}(a,\omega) u(x_i(a,\omega)) \geq \epsilon
\sum_{a\in A_j} \max_{\hat{a}\in A_i} \sum_{\omega\in\Omega} P_{j}(a,\omega) u(x_i(\hat{a},\omega))
\end{eqnarray}
Clearly, NIS passes if and only if IEI is 1, and lower values of IEI indicate that a higher fraction of expected utility has to be shaved off of the RHS to make NIS pass. This can be interpreted as the utility that has to be shaved from every state on the RHS to make NIS pass.%
\footnote{
This is reminiscent of the $\epsilon$-contour set of \textcite{echenique2024individual}, which is all acts that are a $\frac{\epsilon}{1-\epsilon}\%$ improvement over another act.
}

The closest measure to IEI is the Critical Cost Efficiency Index (CCEI) (\cite{afriat1973system}), which measures the budget set reduction necessary for the Generalized Axiom
of Revealed Preference (GARP) to be satisfied for deterministic choice. Like IDI and IEI, the CCEI measures the largest violation in a set of choice data.

\subsection{Statistical Testing}\label{sec:testing}

NIS is a property of idealized state-dependent stochastic choice data \( P \), which cannot be directly observed in experimental data. Instead, experimental data yields an estimator $\hat{P}$ for $P$.
This creates the need for statistical testing.

A number of methods have been used to evaluate the statistical significance of NIAS and NIAC. For example, in \textcite{dewan2020estimating}, NIAS is tested using bootstrapping. If 
no more than 5\% of samples for each action for a given subject fail, then that subject fails to reject NIAS. In \textcite{dewan2020estimating}, NIAC is tested using regression. They run a linear weighted least squares regression of correctness on incentive level and then perform a one-sided \(t\)-test on the coefficient of the incentive level.

This paper evaluates significance using the Wald test. This test calculates the distance between our estimated parameter values \( P(a, \omega) \) across actions and states, and the set of values that satisfy the null hypothesis of NIS, and then compares it with a \(\chi^2\) distribution to determine the \(p\)-value. A key step in the Wald test is calculating the covariance matrix of the difference between the right-hand side and the left-hand side of the NIS inequality, which is 
\begin{equation}
\label{eq:test}
    \sum_{a \in A_i} \sum_{\omega \in \Omega} P_{i}(a, \omega) u(x_i (a, \omega)) - 
    \sum_{a \in A_j} \max_{\hat{a} \in A_i} \sum_{\omega \in \Omega} P_{j}(a, \omega) u(x_i (\hat{a}, \omega)).
\end{equation}
This is achieved using the Delta Method. In Experiment 1.2 of \textcite{dean2023experimental}, the Delta Method is applied to test NIAC. This approach assumes only the asymptotic normality of the estimators. 
For example, in the common case where actions are binary, we observe \(\hat{P}(a \mid \omega)\), which represents a draw from a binomial distribution.
This binomial distribution has a mean of \( P(a \mid \omega) \), with the number of trials determined by the frequency of state \(\omega\).
With a large number of trials, this distribution is approximately normal, validating the assumption.
Additionally, the Delta Method relies on the first-order approximation of the Taylor expansion of expression \eqref{eq:test}. However, since in our case the expression is a linear function of \(\hat{P}(a \mid \omega)\), the first-order approximation of the Taylor expansion is the exact value. This is another reason the Delta Method is especially suitable for our case.

In both \textcite{dewan2020estimating} and \textcite{dean2023experimental}, the prizes, states, and action spaces are all binary and the states are ex ante equally likely.\footnote{In \textcite{caplin2020rational}, the state space and action space have size 5. The expression is slightly longer but follows the same principle.} In that case, the NIS expression \eqref{eq:test} reduces to
\[
P_{i}(a_1 \mid \omega_1) + P_{i}(a_2 \mid \omega_2) - 
\max_{\hat{a} \in \{a_1, a_2\}} \big\{P_{j}(\hat{a} \mid \omega_1) + (1 - P_{j}(\hat{a} \mid \omega_2))\big\} \geq 0
\]
for all \(i, j \in D \). 
This implies
\begin{equation}
\label{eq:nis-bin}
P_{i}(a_1 \mid \omega_1) + P_{i}(a_2 \mid \omega_2) - 
\max_{\hat{a} \in \{a_1, a_2\}} \big\{P_{j}(\hat{a} \mid \omega_1) + (1 - P_{j}(\hat{a} \mid \omega_2))\big\} = 0
\end{equation}
for all \(i \neq j \in D\), and
\begin{equation}
\label{eq:nias-bin}
P_{i}(a_1 \mid \omega_1) + P_{i}(a_2 \mid \omega_2) - 
P_{i}(a_1 \mid \omega_2) - P_{i}(a_2 \mid \omega_1) \geq 0
\end{equation}
for all \(i \in D \).
As discussed in Section \ref{sec:nias}, the latter condition \eqref{eq:nias-bin} within decision problem is equivalent to NIAS in the restricted setting.
Therefore, assuming that NIAS is satisfied and plugging into the preceding equation \eqref{eq:nis-bin} yields the implication that: 
\begin{equation}
    \label{eq:nis-bin-sub}
    P_i (a_1 | \omega_1) + P_i (a_2 | \omega_2) 
    = 
    P_j (a_1 | \omega_1) + P_j (a_2 | \omega_2) 
\end{equation}
for all $i,j \in D$. Equation \eqref{eq:nis-bin-sub} represents a subset of the NIS inequalities in our restricted setting, taking NIAS as given. 
Intuitively, it requires that accuracy be independent of incentives.%
\footnote{
Intuitively, this is a two-sided strengthening of an analogous one-sided implication of NIAC in (\cite{dean2023experimental}, Sec 3.C), namely that subjects become no less accurate as incentives increase. 
}
This is easily tested using an off-the-shelf Wald test for multivariate equality. 
If this relaxed test is rejected, then so is the more stringent NIS condition nesting NIAS. 
Otherwise, it is possible to test the full set of NIS conditions \eqref{eq:nis-bin} and \eqref{eq:nias-bin} using a 
a Wald test for mixed joint hypotheses with equalities and inequalities (\cite{kodde1986wald}).
In what follows, we focus on the simplified implication \eqref{eq:nis-bin-sub}, given that NIAS is typically satisfied pointwise.%
\footnote{
A single exception is the Angles experiment of \textcite{dewan2020estimating} with a \$10 incentive. 
}

\section{Existing Experiments}\label{sec:results_existing}

In this section, we re-examine the data from existing experiments of \textcite{dean2023experimental, caplin2020rational, dewan2020estimating}. Table \ref{tab:summaryresults} provides an overview of the analysis.

\begin{table}[!htbp] 
\centering
\begin{tabular}{llcccccc}
\hline\hline
Experiment & \# of  & NIAS & NIAC & NIS & NIS & IDI &  IEI\\
& incentive  & \multicolumn{2}{c}{point} &rejected & joint & \\
& levels & \multicolumn{2}{c}{estimate} & ($\alpha=0.05$)  & p-value \\
\hline
DN23 1.2 & 4 & Pass & Pass  & Yes & $<0.01$ & 0.17 & 0.89 \\
DN23 2.2 & 2 & Pass & Pass  & No & 0.38  & 0.02 & 0.98 \\
CCLN & 6 & Pass & Fail  & Yes & $<0.01$ & 0.07 & 0.91 \\
CCLN Difficulty 1 & 6 & Pass & Fail & Yes & $<0.01$ & 0.05 & 0.92   \\
CCLN Difficulty 2 & 6 & Pass & Fail & Yes & $<0.01$ & 0.08 & 0.89  \\
CCLN Difficulty 3 & 6 & Pass & Fail & Yes & $<0.01$ & 0.07 & 0.90  \\
CCLN Difficulty 6 & 6 & Pass & Fail & Yes & $<0.01$ & 0.08 & 0.90  \\
DN20 Dots & 4 & Pass & Pass & Yes & $<0.01$  & 0.24 & 0.67 \\
DN20 Dots \$10 & 4 & Pass & Pass & Yes & $<0.01$ & 0.23 & 0.67  \\
DN20 Dots \$20 & 4 & Pass & Pass & Yes & $<0.01$ & 0.25 & 0.66  \\
DN20 Angles & 4 & Pass & Fail & No & 0.54 & 0.02 & 0.96 \\
DN20 Angles \$10 & 4 & Fail & Fail & No & 0.53 & 0.03 & 0.93 \\
DN20 Angles \$20 & 4 & Pass & Fail & No & 0.86 & 0.02 & 0.96  \\
\hline
\end{tabular}
\caption{Summary of results for aggregate data from all experiments.
}
\label{tab:summaryresults}
\end{table}

\subsection{DN23: Experiments 1.2 and 2.2.}

\subsubsection{Design}

As described in Section \ref{sec:preliminaries}, participants in Experiment 1.2 of DN23 face a choice between two actions ($a$ or $b$) with two equally likely states (represented by 49 black and 51 red balls (R) or 49 red and 51 black balls (B)). When the state is R (more red balls), the ``correct'' action is $a$, and when the state if B (more black balls), the correct action is $b$. In our notation, this is analogous to an incentive to choose actions to match the state, i.e. choosing $a_k$ when the state is $\omega_k$ for $k = R,B$. For a randomly selected choice, participants are given probability points (for a prize of \$40) if they chose the correct action and the number of probability points varies across decision problems: either 5, 40, 70, or 95 points. Fifty-two participants face 50 repetitions of each of these four decision problems. Experiment 2.2 has a similar design, but the 55 subjects face seven equations instead of dots with two states of the world being either three or four of the seven equations being correct. The incentive structure remained broadly the same with only the 5 probability points payment and the 95 probability points payment. 

\subsubsection{Results}

As can be gleaned from the joint distribution matrices presented in Section \ref{sec:preliminaries}, subjects improve their accuracy in the direction of increasing incentives, i.e., using the higher incentive-level information structure for a lower incentive-level problem. Thus, in six out of sixteen binary comparisons, specifically, in the direction of increasing incentives, we find an improving switch of attention. 
Table \ref{tab:resultsexp12dn23} summarizes these results using the aggregate data from DN23 Experiment 1.2, with $p$-values corresponding to the one-sided null hypothesis that the NIS RHS is no greater than the LHS for each incentive-level switch.

\begin{table}[!htbp] 
\centering 
\begin{tabular}{@{\extracolsep{5pt}} cccccc} 
\\[-1.8ex]\hline 
\hline \\[-1.8ex] 
\multicolumn{2}{c}{Incentive level} &  &  &  & \\
Lower & Higher & NIS LHS & NIS RHS & NIS inequality fails? & p-value\\ 
\hline \\[-1.8ex] 
$5$ & $40$ & $3.34$ & $3.56$ & Yes & 0.03\\
$5$ & $70$ & $3.34$ & $3.61$ & Yes & 0.02\\
$5$ & $95$ & $3.34$ & $3.77$ & Yes & $<0.01$\\
$40$ & $70$ & $28.46$ & $28.84$ & Yes & 0.33\\
$40$ & $95$ & $28.46$ & $30.12$ & Yes & 0.05\\ 
$70$ & $95$ & $50.47$ & $52.72$ & Yes & 0.11\\ 
\hline \\[-1.8ex] 
\end{tabular} 
  \caption{NIS inequalities in direction of increasing incentives (aggregate data from experiment 1.2 of DN23)}
  \label{tab:resultsexp12dn23} 
\end{table}

If we take the RHS to be the best possible result that can be achieved with the subject's current level of learning and LHS to be the actual level achieved, the difference between them can be understood as an Improvability Index (IDI), as explained in Section \ref{sec:improv}. Our measure of IDI is the maximum improvement over all decision problems, normalized by the best possible outcome in each decision problem. 
The largest normalized improvement relative to baseline comes from the test comparing the performance at the lowest level of incentives (5) with the learning structure from the highest level of incentives (95). The improvement is 0.43 probability points, which corresponds to an increase in 8.6\% of the maximum that could be achieved in that decision problem.



The results from Experiment 2.2 of DN23 suggest the same pattern: the test in the direction of increasing incentives fails NIS.
Yet, the IDI is only 1.6\% -- in other words, in this experiment the improvement gain is not dramatic. 

\begin{table}[!htbp] \centering 
\begin{tabular}{@{\extracolsep{5pt}} cccccc} 
\\[-1.8ex]\hline 
\hline \\[-1.8ex] 
DP1 & DP2 & LHS & RHS & NIS inequality fails? & p-value\\ 
\hline \\[-1.8ex] 
$5$ & $95$ & $4.13$ & $4.21$ & Yes & 0.19\\ 
\hline \\[-1.8ex] 
\end{tabular} 
  \caption{NIS inequalities in direction of increasing incentives (aggregate data from experiment 2.2 of DN23)} 
  \label{tab:resultsexp22dn23} 
\end{table}

\subsection{CCLN}

\subsubsection{Design}

In this experimental task, subjects are shown 24 geometric objects at once. Each one of these objects is a polygon that has either 7, 8, 9 or 10 sides. Subjects are asked to make a binary decision indicating whether they believe that there are more 7-sided polygons (action "heptagon") or 9-sided polygons (action "nonagon"). 
Again, in our notation, this is analogous to an incentive to choose the action to match the state, i.e. choosing $a_k$ when the state is $\omega_k$ for $k = 7,9$.
The number of 7- and 9-sided polygons thus determines the difficulty level of the task: the smaller the difference, the harder the task. \textcite{caplin2020rational} (CCLN hereafter) vary the difference between participants to be either 1, 2, 3, or 6 (the difficulty level is fixed for a given participant). The 8- and 10-sided polygons are thus decoys. Subjects who make a correct decision in a randomly chosen round are rewarded with probability points of winning \$10 -- they receive either 0, 1, 2, 4, 8, 16 or 32 points. To ensure credible probabilistic rewards on MTurk, subjects stopped a computer's built-in clock, with the last two digits providing a uniform random draw; if this number was below their earned probability points score-based threshold minus 100 (e.g., drawing the number 67 with a score of 172), they won the prize (72\% probability in our example), otherwise, they received nothing. Each subject faces 40 rounds in total: 8 at 0 points, 8 at 1 point, 8 at 2 points, 6 at 4 points, 5 at 8 points, 3 at 16 points, and 2 at 32 points. Given that NIS makes no predictions about optimal actions when there are 0 points, we exclude choices under this incentive level.

\subsubsection{Results}

\begin{table}[!htbp] \centering 
 \begin{tabular}{@{\extracolsep{5pt}} cccccc} 
 \hline \hline \\[-1.8ex] 
 DP1 & DP2 & LHS &  RHS & NIS inequality fail? & p-value \\
 \hline
 1 & 2 & 0.63 & 0.66 & Yes & $<0.01$ \\
1 & 4 & 0.63 & 0.66 & Yes &  $<0.01$ \\
1 & 8 & 0.63 & 0.66 & Yes & $<0.01$\\
1 & 16 & 0.63 & 0.69 & Yes & $<0.01$ \\
1 & 32 & 0.63 & 0.68 & Yes &  $<0.01$\\
2 & 4 & 1.31 & 1.32 & Yes & 0.33 \\
2 & 8 & 1.31 & 1.32 & Yes &  0.37 \\
2 & 16 & 1.31 & 1.38 & Yes &  $<0.01$\\
2 & 32 & 1.31 & 1.36 & Yes & 0.03\\
4 & 8 & 2.64 & 2.64 & No & 0.48\\
4 & 16 & 2.64 & 2.76 & Yes &  $<0.01$\\
4 & 32 & 2.64 & 2.72 & Yes &  0.06\\
8 & 16 & 5.27 & 5.52 & Yes & $<0.01$ \\
8 & 32 & 5.27 & 5.44 & Yes &  0.05\\
16 & 32 & 11.05 & 10.87 & No & 0.21 \\
\hline
\end{tabular}
   \caption{NIS inequalities in direction of increasing incentives (aggregate data pooled across difficulty levels in CCLN)} 
   \label{tab:nis_ccln} 
\end{table}

Table \ref{tab:nis_ccln} provides the results aggregating across difficulty levels for the task in CCLN.
In Appendix \ref{apx:ccln}, Tables \ref{tab:nis_ccln_lvl1}, \ref{tab:nis_ccln_lvl2}, \ref{tab:nis_ccln_lvl3} and \ref{tab:nis_ccln_lvl6} provide the results disaggregated by task difficulty level. The NIS test outcomes follow the general pattern we saw previously in DN23: the test generally fails in the direction of increasing incentives (with a joint p-value $< 0.0001$).

\subsection{Dewan and Neligh}

\subsubsection{Design}

\textcite{dewan2020estimating} (DN20 hereafter) use two perceptual tasks with a similar structure. In one version of the task, subjects are shown a random arrangement of dots on the screen and are asked to determine the correct number of dots, between 38 and 42 inclusive, with each number being equally likely. The second task involves an identification of the degrees of an angle presented on the screen -- either 35, 40, 45, 50 or 55 degrees. Thus, both of these decision problems had five possible actions and five possible states. 
Again, in our notation, this is analogous to an incentive to choose the action to match the state.
Each subject was exposed to 100 variations of each task with varying rewards. As is standard, the reward for both tasks was provided in the form of probability points ranging from 1 to 100 for a prize of \$10 in one group of sessions (for 41 subjects) or \$20 in another group of sessions (for 40 subjects) for a total of 8 sessions and 81 subjects. Experimental earnings were based on two randomly selected tasks—one from each half of the experiment—where only correct answers were rewarded, and the incentive level of each selected task determined the probability of winning a monetary prize. Since each subject performed each task at a given incentive level only once, we group the observations at the incentive quartile level. 

\subsubsection{Results}

Aggregate results from the dots task in DN20 adhere to the expected pattern: the six tests in the direction of the increasing incentives, predictably, do not pass the NIS condition. Among the analyzed experiments, the dots task also ranks high on the IDI scale. The largest relative improvement occurs when switching from using the learning attributed to the 1st quartile of incentives (lowest) to the 4th quartile of incentives (highest) and constitutes an aggregate improvement of 23.6 \%. 

\begin{table}[!htbp] 
\centering 
\begin{tabular}{@{\extracolsep{5pt}} cccccc} 
\\[-1.8ex]\hline 
\hline \\[-1.8ex] 
DP 1 (incentive  & DP 2 (incentive  & LHS & RHS & NIS inequality fail? & p-value\\ 
quartile) & quartile) & & & \\
\hline \\[-1.8ex] 
1st & 2nd &$6.18$ & $7.05$ & Yes & {$<0.01$}\\ 
1st & 3rd & $6.18$ & $8.27$ & Yes & {$<0.01$}\\ 
1st & 4th & $6.18$ & $9.25$ & Yes & {$<0.01$} \\ 
2nd & 3rd & $20.62$ & $24.16$ & Yes & {$<0.01$}\\ 
2nd & 4th & $20.62$ & $27.05$ & Yes & {$<0.01$}\\ 
3rd & 4th &$40.05$ & $44.84$ & Yes & {$<0.01$}\\ 
\hline \\[-1.8ex] 
\end{tabular} 
  \caption{NIS inequalities in direction of increasing incentives (aggregate data pooled across prize sizes from the Dot task of DN20)} 
  \label{tab:resultsdotsdn20} 
\end{table}

However, the angles task, a perceptually more difficult task that cannot be verified through effort, shows less responsiveness to the incentives -- the joint probability matrices have little to no change in the direction of increasing incentives. The results of the NIS tests reflect this difficulty: in the direction of increasing incentives, three out of six tests show that participants could not have improved significantly by using a learning structure from the higher incentive tasks. The IDI results also confirm this: while the greatest possible improvement here still comes in the direction of increasing incentives, the magnitude is close to 2\%. 

Thus, NIS is a useful condition for demarcating the kinds of problems in which the returns to cognitive effort are negligible (relative to the costs). 
An interesting avenue for future work would be to verify the condition in settings where incentives may not be monotonically increasing. 

\begin{table}[!htbp] 
\centering 
\begin{tabular}{@{\extracolsep{5pt}} cccccc} 
\\[-1.8ex]\hline 
\hline \\[-1.8ex] 
DP 1 (incentive  & DP 2 (incentive  & LHS & RHS & NIS inequality fail? & p-value\\ 
quartile) & quartile) & & & \\
\hline \\[-1.8ex] 
1st & 2nd & $5.84$ & $5.69$ & No & {0.22}\\ 
    1st & 3rd & $5.84$ & $5.94$ & Yes & {0.29}\\ 
    1st & 4th & $5.84$ & $5.74$ & No & {0.32}\\ 
2nd & 3rd & $16.62$ & $17.37$ & Yes & {0.08}\\
2nd & 4th & $16.62$ & $16.79$ & Yes & {0.36}\\ 
3rd & 4th & $28.80$ & $27.83$ & No & {0.16}\\ 
\hline \\[-1.8ex] 
\end{tabular} 
  \caption{Aggregate NIS test results of the Angle task of DN20} 
  \label{tab:resultsanglesdn20} 
\end{table}

\newpage
\printbibliography

\appendix
\section{Proof of Theorem \ref{thm:ccr}}
\label{apx:proof}

\begin{proof}

Two established properties of the revealed experiment and action strategy defined in \eqref{eq:post-revealed}, \eqref{eq:exp-revealed}, and \eqref{eq:action-revealed} are key: 
\begin{enumerate}
    \item \textcite{caplin2015testable} show that revealed experiment $Q_i^P$ together with the implied mixed action strategies $q_i^P$ generate the data:
    \begin{equation*}
        P_{(Q_i^P,q_i^P)}= P_{i}
    \end{equation*}
    \item \textcite{caplin2015revealed} show that $Q_i^P$ is uniquely the least Blackwell informative experiment that generates the data.
\end{enumerate}

The first point to note is that if a CCR exists, then there exists a CCR in which only the revealed strategies are feasible, $\mathcal{Q}^*\equiv\cup_i Q_i^P$, and in which the revealed strategies are optimal for the corresponding decision problem. 
To see this suppose that a CCR exists and trim the feasible set of experiments to a set of size no more than the number of decision problems by associating with each decision problem $i$ an optimal experiment $Q_i$ with the defining characteristic of a CCR,
\begin{equation}\label{eqn:NISCCR}
  G(Q_i|A_i,x_i)\geq G(Q_j|A_i,x_i)
\end{equation}
for all all $i,j \in D$.

By \textcite{caplin2024modeling}, any form of learning that optimally generates the data is an \textit{optimality preserving spread} of the revealed experiment. Hence replacing $Q_i$, the experiment that is chosen in decision problem $i$ (given choice set $A_i$ and prize specification $x_i$) in the given CCR, with $Q_i^P$ leaves expected utility unchanged:
\begin{eqnarray*}
     G(Q_i|A_i,x_i)&=& G(Q_i^P|A_i,x_i). 
\end{eqnarray*} 

Next, \textcite{caplin2015revealed} show that the revealed information structure is uniquely the least Blackwell informative experiment that generates the data. As a result, for any decision problem, the value of a revealed information structure $Q_i^P$ is no more than its corresponding original information structure $Q_i$. That is for all $j\neq i$:
\begin{eqnarray*}
 G(Q_j|A_i,x_i)\geq G(Q_j^P|A_i,x_i).
\end{eqnarray*} 
We conclude that \eqref{eqn:revealed} holds, i.e., 
\begin{equation*}
  G(Q_i^P|A_i,x_i)\geq G(Q_j^P|A_i,x_i),
\end{equation*}
for all $i,j \in D$, and hence that existence of a CCR implies existence with feasible learning set $\mathcal{Q}^*\equiv\cup_i Q_i^P$.

To complete the proof requires us now to show that 
equation (\ref{eqn:revealed}) is equivalent to $u$ satisfying NIS \eqref{eqn:NIS} for $P$. 
To begin, we can replace the RHS in \eqref{eqn:NIS} with that in \eqref{eqn:revealed} because:
\begin{align*}
    \sum_{a\in A_j} \max_{\hat{a}\in A_i} \sum_{\omega\in\Omega} P_{j}(a,\omega) u(x_i(\hat{a},\omega))
    & \stackrel{(1)}{=} 
    \sum_{a\in A_j} \max_{\hat{a}\in A_i} \sum_{\omega\in\Omega} P_{j}(a) \gamma_j^a (\omega) u(x_i(\hat{a},\omega)) \\ 
    & \stackrel{(2)}{=} 
    \sum_{a\in A_j} P_{j}(a) \max_{\hat{a}\in A_i} \sum_{\omega\in\Omega} \gamma_j^a (\omega) u(x_i(\hat{a},\omega)) \\ 
    & \stackrel{(3)}{=} 
    \sum_{\gamma \in \Gamma (Q_j^P)} Q_j^P (\gamma) \max_{\hat{a}\in A_i} \sum_{\omega\in\Omega} \gamma (\omega) u(x_i(\hat{a},\omega)) \\
    & \stackrel{(4)}{=}
    \sum_{\gamma \in \Gamma (Q_j^P)} Q_j^P (\gamma) \max_{\hat{a}\in A_i} \sum_{\omega\in\Omega} U (\hat{a} | \gamma, x_i) \\ 
    & \stackrel{(5)}{=}
    G( Q_j^P | A_i, x_i)
\end{align*}
where the first equality follows by definition \eqref{eq:post-revealed} of revealed posteriors $\gamma_j^a$, the second by rearrangement, the third by collecting actions $\{a \in A_j : \gamma_j^a  = \gamma \}$ and the definition \eqref{eq:exp-revealed} of the revealed experiment $Q_j^P$, and the fourth by definition of posterior expected utility \eqref{eq:posterior-U}. Finally the fifth equality follows because (by the fourth equality), the RHS in \eqref{eqn:NIS} is a restriction of \eqref{eq:G} to pure strategies $q (a|\gamma) \in \{0,1\}$, which establishes an upper bound; and yet, an optimal solution to linear program \eqref{eq:G} in pure strategies always exists, which establishes equality. 
For the LHS, we note further that: 
\[
    \sum_{a\in A_i} \max_{\hat{a}\in A_i} \sum_{\omega\in\Omega} P_{i}(a,\omega) u(x_i(\hat{a},\omega))
    = 
    \sum_{a\in A_i} \sum_{\omega\in\Omega} P_{i}(a,\omega) u(x_i(a,\omega))
\]
by \eqref{eq:NIAS-equiv}, i.e., NIAS or NIS restricted to case $i = j$. 
Substitution on both sides of equation (\ref{eqn:revealed}) then establishes the validity of equation (\ref{eqn:NIS}) and with it the proof.
\end{proof}

\section{Disaggregated Experiment Results}

\subsection{CCLN, by Difficulty Level}
\label{apx:ccln}

 \begin{table}[H] \centering 
 \begin{tabular}{@{\extracolsep{5pt}} cccccc} 
\\[-1.8ex]\hline 
\hline \\[-1.8ex] 
\multicolumn{2}{c}{Incentive level} &  &  &  & \\
Lower & Higher & NIS LHS & NIS RHS & NIS inequality fail? & p-value\\ 
\hline \\[-1.8ex] 
$1$ & $2$ & $0.58$ & $0.61$ & Yes & 0.02\\ 
 $1$ & $4$ & $0.58$ & $0.61$ & Yes & 0.03\\
 $1$ & $8$ & $0.58$ & $0.63$ & Yes & $<0.01$\\ 
 $1$ & $16$ & $0.58$ & $0.63$ & Yes & $<0.01$\\
 $1$ & $32$ & $0.58$ & $0.59$ & Yes & 0.31\\ 
 $2$ & $4$ & $1.22$ & $1.22$ & Yes & 0.48\\ 
 $2$ & $8$ & $1.22$ & $1.25$ & Yes & 0.20\\
 $2$ & $16$ & $1.22$ & $1.26$ & Yes & 0.20\\ 
 $2$ & $32$ & $1.22$ & $1.18$ & No & 0.22\\ 
 $4$ & $8$ & $2.45$ & $2.50$ & Yes & 0.22\\ 
 $4$ & $16$ & $2.45$ & $2.52$ & Yes & 0.19\\ 
 $4$ & $32$ & $2.45$ & $2.37$ & Yes & 0.23\\ 
 $8$ & $16$ & $5.01$ & $5.03$ & Yes & 0.45\\ 
 $8$ & $32$ & $5.01$ & $4.73$ & No  & 0.09\\
 $16$ & $32$ & $10.06$ & $9.47$ & No & 0.08\\ 
  \hline 
 \end{tabular}
    \caption{NIS inequalities in direction of increasing incentives (aggregate data from difficulty level 1 in CCLN)} 
   \label{tab:nis_ccln_lvl1} 
\end{table}

\begin{table}[H] \centering 
 \begin{tabular}{@{\extracolsep{5pt}} cccccc} 
 \hline \hline \\[-1.8ex] 
 DP1 & DP2 & LHS &  RHS & NIS inequality fail? & p-value \\
 \hline
 $1$ & $2$ & $0.60$ & $0.64$ & Yes & {$<0.01$}\\ 
 $1$ & $4$ & $0.60$ & $0.65$ & Yes & {$<0.01$} \\
 $1$ & $8$ & $0.60$ & $0.64$ & Yes & {$<0.01$} \\ 
 $1$ & $16$ & $0.60$ & $0.68$ & Yes & {$<0.01$}\\ 
 $1$ & $32$ & $0.60$ & $0.66$ & Yes & {$<0.01$}\\ 
 $2$ & $4$ & $1.29$ & $1.31$ & Yes & {0.28}\\ 
 $2$ & $8$ & $1.29$ & $1.29$ & Yes &  {0.49}\\
 $2$ & $16$ & $1.29$ & $1.35$ & Yes  & {0.05}\\ 
 $2$ & $32$ & $1.29$ & $1.32$ & Yes  & {0.22}\\ 
 $4$ & $8$ & $2.61$ & $2.57$ & No &  {0.29} \\ 
 $4$ & $16$ & $2.61$ & $2.71$ & Yes & {0.14}\\
 $4$ & $32$ & $2.61$ & $2.65$ & Yes & {0.36}\\ 
 $8$ & $16$ & $5.15$ & $5.41$ & Yes & {0.04} \\
 $8$ & $32$ & $5.15$ & $5.29$ & Yes  & {0.47}\\ 
 $16$ & $32$ & $10.82$ & $10.56$ & No & {0.24}\\ 
 \hline 
\end{tabular}
   \caption{NIS inequalities in direction of increasing incentives (aggregate data from difficulty level 2 in CCLN)} 
   \label{tab:nis_ccln_lvl2} 
\end{table}

\begin{table}[H] 
\centering 
\begin{tabular}{@{\extracolsep{5pt}} cccccc} 
\hline\hline \\[-1.8ex] 
DP1 & DP2 & LHS & RHS & NIS inequality fail? & p-value \\ 
\hline \\[-1.8ex] 
$1$ & $2$ & $0.62$ & $0.65$ & Yes & {$0.01$} \\
$1$ & $4$ & $0.62$ & $0.64$ & Yes & {$0.09$} \\ 
$1$ & $8$ & $0.62$ & $0.65$ & Yes & {$0.08$} \\ 
$1$ & $16$ & $0.62$ & $0.68$ & Yes & {$<0.01$} \\
$1$ & $32$ & $0.62$ & $0.69$ & Yes & {$<0.01$} \\ 
$2$ & $4$ & $1.30$ & $1.28$ & Yes &{$0.24$} \\ 
$2$ & $8$ & $1.30$ & $1.29$ & Yes & {$0.37$} \\
$2$ & $16$ & $1.30$ & $1.37$ & Yes & {$0.07$} \\
$2$ & $32$ & $1.30$ & $1.37$ & Yes & {$0.10$} \\ 
$4$ & $8$ & $2.56$ & $2.58$ & Yes & {$0.42$} \\
$4$ & $16$ & $2.56$ & $2.73$ & Yes & {$0.02$} \\ 
$4$ & $32$ & $2.56$ & $2.74$ & Yes & {$0.04$} \\ 
$8$ & $16$ & $5.16$ & $5.46$ & Yes & {$0.06$} \\ 
$8$ & $32$ & $5.16$ & $5.48$ & Yes & {$0.05$} \\ 
$16$ & $32$ & $10.93$ & $10.96$ & Yes & {$0.47$} \\ 
\hline 
\end{tabular} 
\caption{NIS inequalities in direction of increasing incentives (aggregate data from difficulty level 3 in CCLN)} 
\label{tab:nis_ccln_lvl3} 
\end{table} 

\begin{table}[H] 
\centering 
\begin{tabular}{@{\extracolsep{5pt}} cccccc} 
\hline \hline \\[-1.8ex] 
DP1 & DP2 & LHS & RHS & NIS inequality fail? & p-value \\ 
\hline \\[-1.8ex] 
$1$ & $2$ & $0.72$ & $0.73$ & Yes & {$0.22$} \\ 
$1$ & $4$ & $0.72$ & $0.75$ & Yes & {$0.04$} \\ 
$1$ & $8$ & $0.72$ & $0.73$ & Yes & {$0.24$} \\ 
$1$ & $16$ & $0.72$ & $0.79$ & Yes & {$<0.01$} \\ 
$1$ & $32$ & $0.72$ & $0.80$ & Yes & {$<0.01$} \\
$2$ & $4$ & $1.46$ & $1.49$ & Yes & {$0.15$} \\ 
$2$ & $8$ & $1.46$ & $1.46$ & Yes & {$0.47$} \\ 
$2$ & $16$ & $1.46$ & $1.57$ & Yes & {$<0.01$} \\ 
$2$ & $32$ & $1.46$ & $1.59$ & Yes & {$<0.01$} \\
$4$ & $8$ & $2.98$ & $2.92$ & No & $0.22 $\\
$4$ & $16$ & $2.98$ & $3.14$ & Yes & {$0.02$} \\ 
$4$ & $32$ & $2.98$ & $3.19$ & Yes & {$0.01$} \\
$8$ & $16$ & $5.84$ & $6.28$ & Yes & {$<0.01$} \\ 
$8$ & $32$ & $5.84$ & $6.37$ & Yes & {$<0.01$} \\
$16$ & $32$ & $12.57$ & $12.75$ & Yes & {$0.32$} \\ 
\hline 
\end{tabular} 
\caption{NIS inequalities in direction of increasing incentives (aggregate data from difficulty level 6 in CCLN)} 
\label{tab:nis_ccln_lvl6} 
\end{table} 

\subsection{DN20 Dots Task, by Incentive Level}
\label{apx:dn20dots}

\begin{table}[H] 
\centering 
\begin{tabular}{@{\extracolsep{5pt}} cccccc} 
\\[-1.8ex]\hline 
\hline \\[-1.8ex] 
DP 1 (incentive  & DP 2 (incentive  & LHS & RHS & NIS inequality fail? & p-value\\ 
quartile) & quartile) & & & \\
\hline \\[-1.8ex] 
1st & 2nd & $6.03$ & $6.94$  & Yes & {$<0.01$}\\ 
1st & 3rd & $6.03$ & $7.82$ & Yes & {$<0.01$}\\ 
1st & 4th & $6.03$ & $8.95$ & Yes & {$<0.01$}\\ 
2nd & 3rd & $20.28$ & $22.86$ & Yes & {$<0.01$}\\ 
2nd & 4th & $20.28$ & $26.16$ & Yes & {$<0.01$}\\ 
3rd & 4th & $37.89$ & $43.37$ & Yes & {$<0.01$}\\ 
\hline \\[-1.8ex] 
\end{tabular}
  \caption{Aggregate NIS test results of the Dot task of DN20 (only \$10 sessions)} 
  \label{tab:resultsdotsdn20_10} 
\end{table}

\begin{table}[H] 
\centering 
\begin{tabular}{@{\extracolsep{5pt}} cccccc} 
\\[-1.8ex]\hline 
\hline \\[-1.8ex] 
DP 1 (incentive  & DP 2 (incentive  & LHS & RHS & NIS inequality fail? & p-value\\ 
quartile) & quartile) & & & \\
\hline \\[-1.8ex] 
1st & 2nd & $6.33$ & $7.16$  & Yes & {$<0.01$}\\ 
1st & 3rd & $6.33$ & $8.74$ & Yes & {$<0.01$}\\
1st & 4th & $6.33$ & $9.57$ & Yes & {$<0.01$}\\ 
2nd & 3rd & $20.92$ & $25.54$ & Yes& {$<0.01$}\\
2nd & 4th & $20.92$ & $27.98$ & Yes & {$<0.01$}\\ 
3rd & 4th & $42.35$ & $46.39$ & Yes & {$<0.01$}\\ 
\hline \\[-1.8ex] 
\end{tabular} 
  \caption{Aggregate NIS test results of the Dot task of DN20 (only \$20 sessions)} 
  \label{tab:resultsdotsdn20_20} 
\end{table}

\subsection{DN20 Angles Task, by Incentive Level}
\label{apx:dn20angles}

\begin{table}[H] 
\centering  
\begin{tabular}{@{\extracolsep{5pt}} cccccc} 
\\[-1.8ex]\hline 
\hline \\[-1.8ex] 
DP 1 (incentive  & DP 2 (incentive  & LHS & RHS & NIS inequality fail? & p-value\\ 
quartile) & quartile) & & & \\
\hline \\[-1.8ex] 
1st & 2nd & $5.64$ & $5.604$  & No & {0.45} \\
1st & 3rd & $5.64$ & $5.91$ & Yes & {0.18}\\
1st & 4th & $5.64$ & $5.68$ & Yes & {0.35}\\ 
2nd & 3rd & $16.38$ & $17.27$ & Yes & {0.13}\\ 
2nd & 4th & $16.38$ & $16.62$ & Yes & {0.37}\\ 
3rd & 4th & $28.63$ & $27.55$ & No & {0.08}\\
\hline \\[-1.8ex] 
\end{tabular} 
  \caption{Aggregate NIS test results of the Angle task of DN20 (\$10 sessions only)} 
  \label{tab:resultsanglesdn20_10}
\end{table}

\begin{table}[H] 
\centering 
\begin{tabular}{@{\extracolsep{5pt}} cccccc} 
\\[-1.8ex]\hline 
\hline \\[-1.8ex] 
DP 1 (incentive  & DP 2 (incentive  & LHS & RHS & NIS inequality fail? & p-value\\ 
quartile) & quartile) & & & \\
\hline \\[-1.8ex] 
1st & 2nd & $6.02$ & $5.80$  & No & {0.20}\\
1st & 3rd & $6.02$ & $5.95$ & No & {0.39}\\
1st & 4th & $6.02$ & $5.99$ & No & {0.43}\\
2nd & 3rd & $16.94$ & $17.38$ & Yes & {0.27}\\
2nd & 4th & $16.94$ & $17.51$ & Yes & {0.24}\\ 
3rd & 4th &$28.81$ & $29.02$ & Yes & {0.45}\\ 
\hline \\[-1.8ex] 
\end{tabular} 
  \caption{Aggregate NIS test results of the Angle task of DN20 (\$20 sessions only)
  } 
  \label{tab:resultsanglesdn20_20} 
\end{table}

\end{document}